\documentclass[aps,prb,preprint,groupedaddress,showpacs]{revtex4-1}
\usepackage{graphicx}% Include figure files
\usepackage{dcolumn}% Align table columns on decimal point
\usepackage{bm}% bold math
\usepackage{amsmath}% Split the equation
\begin{document}

\preprint{APS/123-QED}

\title{Exchange constant and domain wall width in (Ga,Mn)(As,P) films with self-organization of magnetic domains}

\author{S. Haghgoo$^{1,2}$,  M. Cubukcu$^{1,2}$, H. J. von Bardeleben$^{1,2}$, L. Thevenard$^{1,2}$, A. Lema\^{\i}tre$^3$, and C. Gourdon$^{1,2}$}
 %\altaffiliation[Also at ]{}%Lines break automatically or can be forced with \\
\email[e-mail: ]{gourdon@insp.jussieu.fr}
%\homepage[web site: ]{http://www.insp.upmc.fr}
 
\affiliation{
$^1$ Institut des Nanosciences de Paris, CNRS, UMR7588, 140 rue de Lourmel, Paris, F-75015 France\\
$^2$ Universit\'{e} Pierre et Marie Curie, UMR 7588, INSP, Paris, France\\
$^3$ Laboratoire de Photonique et Nanostructures, CNRS, UPR 20, Route de Nozay, Marcoussis, F-91460 France}

\date{\today}% It is always \today, today,

\label{sec:Abstract}

\begin{abstract}
%\section{Abstract}
The incorporation of Phosphorus into (Ga,Mn)As epilayers allows for the tuning of the magnetic easy axis from in-plane to perpendicular-to-plane without the need for a (Ga,In)As template. For perpendicular easy axis, using magneto-optical imaging a self-organized pattern of up- and down-magnetized domains is observed for the first time in a diluted magnetic semiconductor. Combining Kerr microscopy, magnetometry and ferromagnetic resonance spectroscopy, the exchange constant and the domain wall width parameter are obtained as a function of temperature. The former quantifies the effective Mn-Mn ferromagnetic interaction. The latter is a key parameter for domain wall dynamics. The comparison with results obtained for (Ga,Mn)As/(Ga,In)As reveals the improved quality of the (Ga,Mn)As$_{1-y}$P$_y$ layers regarding domain wall pinning, an increase of the domain wall width parameter and of the effective Mn-Mn spin coupling. However, at constant Mn doping, no significant increase of this coupling is found with increasing P concentration in the investigated range. 
% abstract max 100 mots pour APL
\end{abstract}

%Magnetic semiconductors, Domain walls and domain structure, Magnetic properties of monolayers and thin films  

\label{sec:Introduction}

\maketitle
%\section{Introduction}
Diluted ferromagnetic (FM) (III,Mn)V semiconductors are the subject of intensive research in view of potential applications in the field of spintronics~\cite{Jungwirth2006,Awschalom2007}. Several proposals for magnetization manipulation using either light-induced magnetization switching between different magnetic easy axes~\cite{Astakhov2005,Munekata2005} or magnetic domain wall (DW) motion~\cite{Tang2006,Yamanouchi2006,Dourlat2008,Adam2009a} have emerged. (Ga,Mn)As layers and microtracks with perpendicular magnetic anisotropy are well suited for the investigation of DW propagation. In such layers the intrinsic flow regime for field-driven DW propagation was demonstrated for the first time in a diluted FM semiconductor~\cite{Dourlat2008}.
\par
One of the  properties of this versatile FM semiconductor is the strong dependence of its magnetic anisotropy on epitaxial strain~\cite{Dietl2001,Jungwirth2006,Thevenard2006,Dourlat2007}. This dependence results from the carrier-mediated origin of the ferromagnetism. It arises from the competition between the Mn-induced giant Zeeman splitting and the strain-induced splitting of the valence bands.
Recently, it was shown that introducing a few percent of Phosphorus in substitution of Arsenic in (Ga,Mn)As layers grown on GaAs provides an efficient tuning of the epitaxial strain, from compressive to tensile, due to the smaller ionic radius of Phosphorus~\cite {Cubukcu2010,Lemaitre2008}. As expected theoretically, the magnetic easy axis switches from in-plane to out-of-plane orientation upon increasing P concentration. This eliminates the need for a highly mismatched (Ga,In)As buffer layer for realizing out-of-plane anisotropy. This buffer layer was shown to introduce a cross-hatch pattern and emerging dislocations which are detrimental to the propagation of magnetic DWs~\cite{Thevenard2006,Dourlat2007}. (Ga,Mn)As$_{1-y}$P$_y$ films grown directly on a GaAs buffer are expected to contain much less defects. In addition to this structural advantage, (Ga,Mn)As$_{1-y}$P$_y$ films are appealing for two aspects~\cite{Masek2007}. Firstly, it is expected that the smaller lattice constant of (Ga,Mn)As$_{1-y}$P$_y$ will hinder Mn incorporation in interstitial sites, where Mn ions act as charge and magnetic moment compensating defects, thereby favoring higher carrier density than in (Ga,Mn)As. Secondly, it was predicted that the decrease of the lattice constant should also lead to an increase of the Mn-hole exchange integral $J_{pd}$ and hence of the Curie temperature.  In turn, this should lead to an increase of the exchange constant $A$ describing the effective exchange interaction between the Mn spins in the framework of the micromagnetic theory. The determination of $A$ yields the DW width parameter defined as $\Delta=\sqrt{A/K_u}$, where $K_u$ is the uniaxial anisotropy constant. $\Delta$ is a key parameter for DW dynamics~\cite{Slonczewski1974}. The Walker velocity, \textit{i.e.} the theoretical critical velocity at the crossover from the steady to the precessional flow regimes in the one-dimensional model, depends on $\Delta$ as $v_W=\gamma\Delta\mu_0 M_s /2$, where $\gamma$ is the electron gyromagnetic ratio and $M_s$ the magnetization. Furthermore, in both the steady and precessional regimes, the DW mobility is proportional to $\Delta$.
\par
In FM layers with perpendicular easy axis one of the methods for the determination of the exchange constant $A$ is based on domain theory, \textit{i.e.} micromagnetic theory applied to self-organized magnetic domains~\cite{Hubert2000,Dietl2001a,Vertesy2003,Gourdon2007}. Self-organization of up- and down- magnetized domains in a periodic array results from the competition between the DW energy and the magnetic energy arising from long-range interaction between domains. Up to now, although an estimation of $A$ could be obtained from the domain structure and hysteresis cycle in (Ga,Mn)As~\cite{Gourdon2007}, a long-range periodic arrangement of magnetic domains in (III,Mn)V FM semiconductors had not yet been obtained.
%\section{Goal}
\par
In this paper we report the observation of self-organized magnetic domains in (Ga,Mn)As$_{1-y}$P$_y$ FM alloys with perpendicular magnetic anisotropy. We determine the exchange constant $A$ and the DW width parameter $\Delta$ from the domain period. An effective exchange coupling constant $J_{MnMn}$ between Manganese spins is then obtained as a function of Phosphorus concentration.%\section{samples}
\label{sec:samples}
\par
The samples were grown by molecular beam epitaxy on GaAs (001) substrates. Details can be found in Ref.~[\onlinecite{Thevenard2006},\onlinecite{Lemaitre2008}]. For (Ga,Mn)As$_{1-y}$P$_y$  samples, the Mn (P) concentration was estimated from  
a reference (Ga,Mn)As ((Ga,As)P) sample, grown under similar conditions, in particular a similar substrate temperature. However the determination of Mn and P concentrations is made difficult by the presence of an unknown concentration of interstitial Mn and As in antisites. In sample D, with no Phosphorus, the (Ga,Mn)As layer was grown in the traditional manner on a Ga$_{1-z}$In$_{z}$As buffer layer (z=0.098) in order to achieve perpendicular anisotropy~\cite{Thevenard2006}. To improve $M_s$, the Curie temperature $T_C$ and the carrier concentration the samples were annealed under N$_2$ atmosphere at 250$^{\circ}$C for 1 hour. For all samples the layer thickness is $d=50$~nm. 
\par
The magnetic domain structure was investigated using polar magneto-optical Kerr (MOKE) microscopy at variable temperature~\cite{Dourlat2007}. The lateral spatial resolution is 0.9 $\mu$m. The light wavelength is 600 (670) nm for the (Ga,Mn)AsP ((Ga,Mn)As) samples. The hysteresis cycle was obtained from the average intensity of MOKE images as a function of the applied field. The temperature dependence of the saturation magnetization $M_s(T)$ and the Curie temperature $T_C$ were obtained from magnetometry using a superconductor quantum interference device (SQUID). The anisotropy constants were determined from ferromagnetic resonance (FMR) spectroscopy. The spectra were analysed using the Smit-Beljers equation and the minimization of the free energy for different alignments of the applied magnetic field~\cite{Smit1955,Cubukcu2010}. 
\par
%\begin{table}[t]
%\caption{\label{table1} (Ga,Mn)As$_{1-y}$P$_y$ and (Ga,Mn)As sample parameters. The lattice mismatch (lm) is defined as $(a_\bot-a_{sub})/a_{sub}$, where $a_\bot(a_{sub})$ is the lattice parameter of the film (substrate) parallel to the growth axis. $\epsilon_{zz}$ is the corresponding calculated strain component. $M_s$ and $Q$ are the magnetization and anisotropy quality factor, respectively, at temperature $T=4$~K. $Q$ slightly increases with temperature.\\}
%\begin{ruledtabular}
%\scriptsize
%\begin{tabular*}{0.2\textwidth}{@{\extracolsep{2.5mm}}c|c|c|c||c||c||c}
%sample&A1&A2&A3&B&C&D\\\hline
%[Mn$_{total}$](\%)&10.4&10.4&10.4&8&7& 7 \\\
%[P](\%)&11.3&8.8&7&8.5& 7&0 \\
%lm (ppm)&-6650&-5120&-4260&-6870&-3580&-10700 \\
%$\epsilon_{zz}$(\%)&-0.3&-0.2&-0.2&-0.3&-0.2&-0.5 \\
%$T_C$~(K) &119&113&139&110&80&125 \\
%$M_s$ (kA m$^{-1}$)&53.8&50.8&53.5&46.2&39.2&38.5\\
%$Q$ &4.35&3.3&3&4.8&3.25&8.5 \\
%\end{tabular*}
%\end{ruledtabular}  
%\end{table}

\par
Table~\ref{table1} summarizes the characteristics of the six samples which were selected in order to compare layers with the same Mn concentration and different P concentrations (A series, C and D), or the same P concentration and different Mn concentrations (A2 and B, A3 and C). Most of the samples show a $T_C$ around 110 K. $T_C$ is not improved significantly in the (Ga,Mn)As$_{1-y}$P$_y$ samples as compared to sample D (no Phosphorus). The concentration of Mn ions participating in the ferromagnetism [Mn$_{eff}$] is determined from the saturation magnetization $M_s$ at $T=4$~K as [$Mn_{eff}$]=$M_s/(N_0g\mu_{B}S)$, where $N_0$ is the density of cation sites in the lattice, $\mu_B$ the Bohr magneton and $S=5/2$. For all samples the ratio [Mn$_{eff}$]/[Mn$_{total}$] is $\approx$0.5, revealing the fact that the incorporation of Mn in interstitial sites is not modified by  alloying with Phosphorus.  
\par
%theory, méthode d'analyse
The method employed to determine the exchange constant and the DW width parameter is based on domain theory for layers with perpendicular easy axis~\cite{Hubert2000}. Self-organization of up- and down-magnetized domains in a periodic stripe array minimizes the total energy consisting of the Zeeman energy, the surface energy of the DWs and the stray field energy. The stripe period $p$ (in units of $d$) in the demagnetized state (zero applied field and zero average magnetization) satisfies the following equation~\cite{Hubert2000}:
\begin{widetext} 
\begin{equation*}
    \lambda_c=\frac{p}{\pi^3 Q} \sum_{n=1}^{\infty} \frac{Sin^2{\left(\frac{n\pi}{2}\right)}\times\left[-pQ-2n\pi(1+Q)+pQ\left(Cosh\left(\chi\right)+\sqrt{1+\frac{1}{Q}} Sinh\left(\chi\right)\right)\right]}{n^{3}\left(\sqrt{1+\frac{1}{Q}}Cosh\left(\chi\right)+Sinh\left(\chi\right)\right)^2}\,
\end{equation*}
\end{widetext}  
where $\chi$ is equal to $\left(2n\pi\sqrt{1+1/Q}\right)/p$. $Q =K_u/(\mu_0 M_s^2/2)$ is the anisotropy quality factor shown in Table~\ref{table1}. Q is smaller in the P incorporated samples than in sample D due to the smaller $K_u$ and larger $M_s$. $\lambda_c$ is Thiele's length defined as $\sigma/\mu_0M_s^2 d$ with $\sigma$ the specific DW energy. Knowing $p$ and $Q$, one determines $\lambda_c$ and hence $\sigma$. Typically, $\lambda_c$ is found to be  of the order of 1. For a Bloch wall $\sigma$ is equal to $4\sqrt{AK_{u}}$. Micromagnetic simulations using OOMMF package~\cite{oommf} show that in our samples with $d$=50~nm DWs are not Bloch walls but twisted walls with N\'{e}el caps. Using the variational calculations of Ref.~[\onlinecite{Hubert1975}] for the DW specific energy of twisted walls we obtain the functional dependence of $\sigma$ on $Q$ and $\Delta$ namely $\sigma= 4 f(Q,\Delta)\Delta K_{u}$ with $f(Q,\Delta)$$\leq$ $1$. For each value of $\sigma$ we obtain the DW width parameter $\Delta$ and hence the exchange constant $A$.
%
%
%\section{Results}
%\begin{figure}[h,t,b]
%\begin{center}
%\includegraphics[width=1\linewidth]{periode-image.eps}
%\end{center}
 %       \caption{(a) Self-organized pattern in the demagnetized state for sample A3 at $T=80$~K. (b) Domain period as a function of temperature for sample A3.}
  %      \label{periode-image}
%\end{figure}
%
%

\begin{figure}[h,t,b]
\begin{center}
\includegraphics[width=1\linewidth]{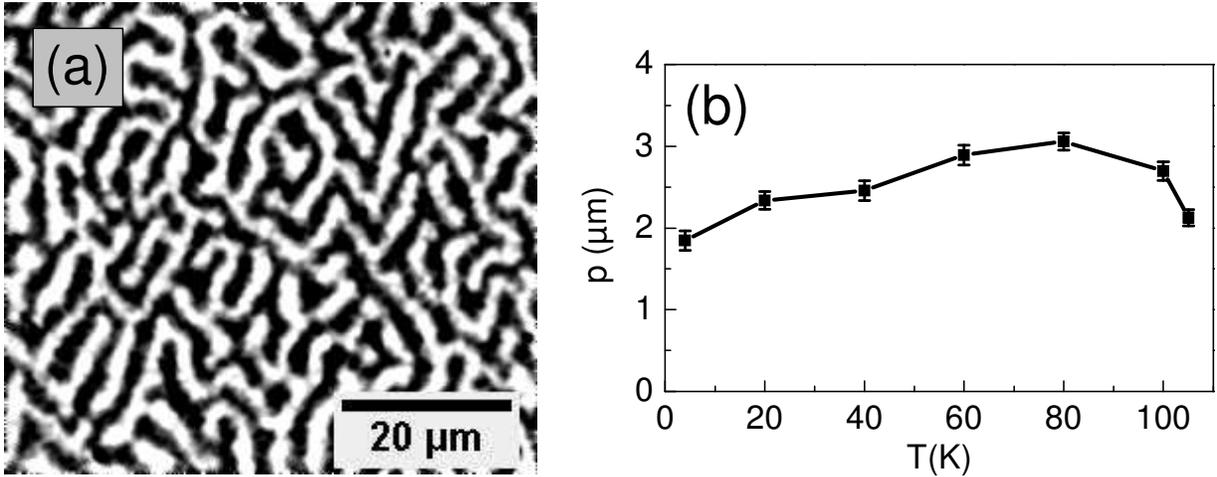}
\end{center}
        \caption{(a) Self-organized pattern in the demagnetized state for sample A3 at $T=80$~K. (b) Domain period as a function of temperature for sample A3.}
       \label{periode-image}
\end{figure}

\begin{figure}[t,h,b]
\begin{center}
        \includegraphics[width=0.98\linewidth]{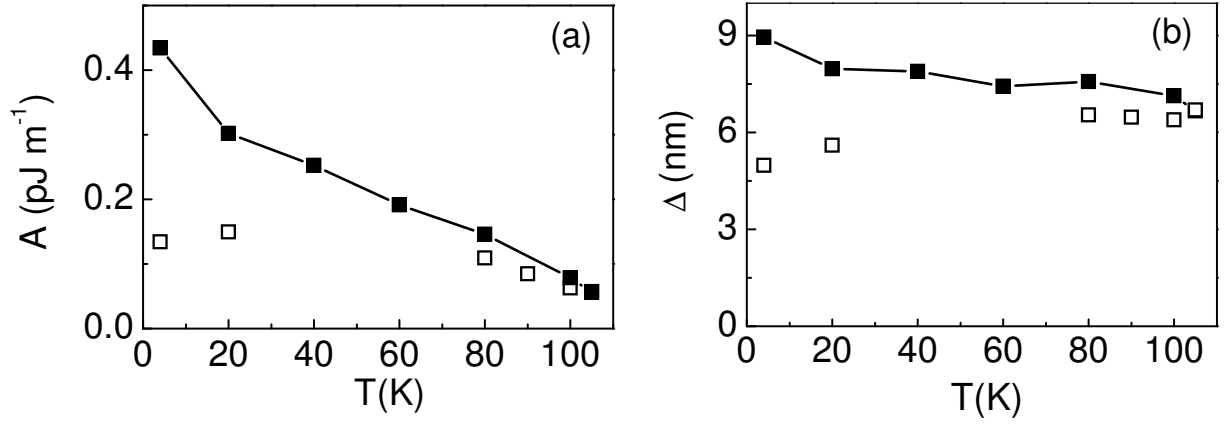}
        \end{center} 
\caption{Exchange constant $A$ (a) and DW width parameter $\Delta$ (b) for sample A3. The filled and open symbols represent the values determined from the period of self-organized magnetic domains and from the hysteresis cycle, respectively.}
      \label{Aanddelta}
\end{figure}

\begin{figure}[h,t,b]
\begin{center}
                \includegraphics[width=0.6\linewidth]{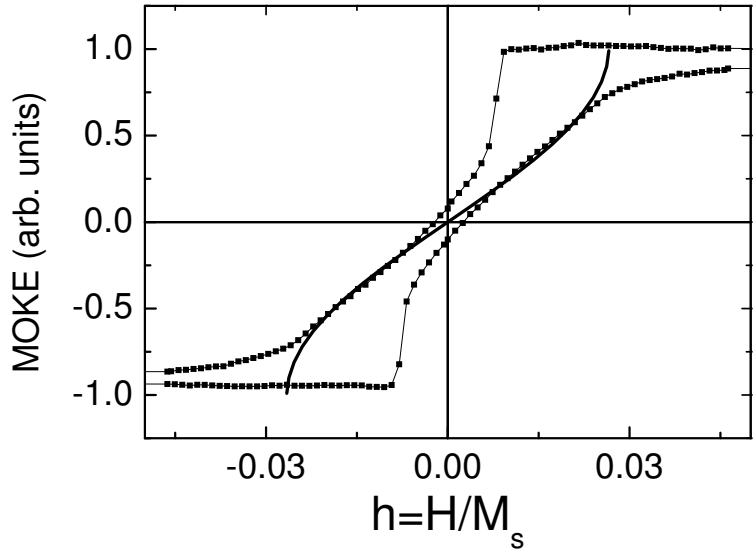}
\end{center}
        \caption{Experimental hysteresis cycle (symbols) for sample A3 (10.4~\% Mn, 7~\% P) at $T=$100~K and calculated $m(h)$ curve for $\lambda_c=0.95$ (solid line).}
     \label{cycle}
\end{figure}

Figure~\ref{periode-image}(a) shows a typical MOKE image of the self-organized domains with up- and down- magnetization (white and black) after a demagnetization process in an alternating magnetic field of decreasing amplitude. The remnant field is $\leq1$~G. This image is obtained on sample A3 at $T=80$~K. Figure~\ref{periode-image}(b) shows the domain period obtained from the MOKE images as a function of the temperature $T$ for this sample. The period is found in the range 2-3 $\mu$m for 4~K$<T<$105~K. The period for 4~K$\leq T\leq $40~K and $T$=105~K is determined from MOKE images with a weak optical contrast. We will discuss below the reliability of the data extracted from these values. Following the procedure described above, Thiele's length $\lambda_c$, the DW specific energy $\sigma$, the DW width parameter $\Delta$, and eventually the exchange constant $A$ are obtained.  $\Delta$ and $A$ are shown in Fig.~\ref{Aanddelta} for sample A3 (filled square symbols) as a function of temperature. $\Delta$ is found in the range 7-9~nm and $A$ in the range $0.1-0.4$ pJ~m$^{-1}$. The decrease of $A$ with temperature mainly reflects the decrease of the magnetization ($A\propto\lambda_c^2 M^4/K_u$). 
%\begin{figure}[t,h,b]
%\begin{center}
%        \includegraphics[width=0.98\linewidth]{Aanddelta.eps}
%        \end{center} 
%\caption{Exchange constant $A$ (a) and DW width parameter $\Delta$ (b) for sample A3. The filled and open symbols represent the values determined from the period of self-organized magnetic domains and from the hysteresis cycle, respectively.}
 %       \label{Aanddelta}
%\end{figure}
%
In order to assess the reliability of the $\Delta$ and $A$ values at low and high temperature, we determine the upper and lower boundaries for these quantities using the method previously developed for (Ga,Mn)As/GaInAs~\cite{Gourdon2007}. The upper boundary for $\lambda_c$ and consequently for $\Delta$ and $A$ is determined from the width of the stripe domains of the minority phase observed close to the saturation field. For sample A3 this upper boundary coincides at all temperatures with the values obtained from the domain period. The lower boundary for $\lambda_c$ (and hence for $\Delta$ and $A$) is obtained from the comparison of the normalized experimental hysteresis cycle with the calculated magnetization curve $m(h)$, where $m$ is the spatially averaged reduced magnetization $\left\langle M\right\rangle/M_s$ and $h$ the reduced applied magnetic field $H/M_s$. The $m(h)$ curve is obtained from the minimization of the free energy of the periodic stripe array. $\lambda_c$ is the only free parameter. An example of experimental hysteresis cycle and calculated $m(h)$ curve is given in Fig.~\ref{cycle}. The $m(h)$ curve which is tangent to the experimental hysteresis cycle provides the lower boundary for $\lambda_c$ and hence for $\Delta$ and $A$. The values obtained by this method are shown by open symbols in Fig.~\ref{Aanddelta}. At high temperature (90-100~K) the lower boundary values coincide with the values determined from the domain period. To summarize, between 4~K and 40~K the DW width parameter and exchange constant are comprised between the values shown by the closed and open symbols of Fig.~2, respectively. Above 40~K $\Delta$ and $A$ are reliably obtained from the domain period (filled symbols).
%
%\begin{figure}[h,t,b]
%\begin{center}
 %               \includegraphics[width=0.6\linewidth]{cycle.eps}
%\end{center}
 %       \caption{Experimental hysteresis cycle (symbols) for sample A3 (10.4~\% Mn, 7~\% P) at $T=$100~K and calculated $m(h)$ curve for $\lambda_c=0.95$ (solid line).}
  %      \label{cycle}
%\end{figure}
%
Similar results are obtained for samples A1 and A2 where the self-organized domain pattern can be achieved. For samples B and C DW pinning by defects prevents the self-organization of domains in regular patterns. Therefore only the lower and upper boundaries of $\Delta$ and $A$ can be obtained, using the method recalled above. For sample C the upper boundary values could not be obtained below 30~K since the stripe domains were not observed. For sample D ((Ga,Mn)As on a (Ga,In)As template) the DW width and exchange constant were obtained from the analysis of the field-driven DW dynamics~\cite{Dourlat2008} in excellent agreement with the estimation from the domain structure and hysteresis cycle~\cite{Gourdon2007}.

\begin{figure}[t,b]
        \includegraphics[width=0.7\linewidth]{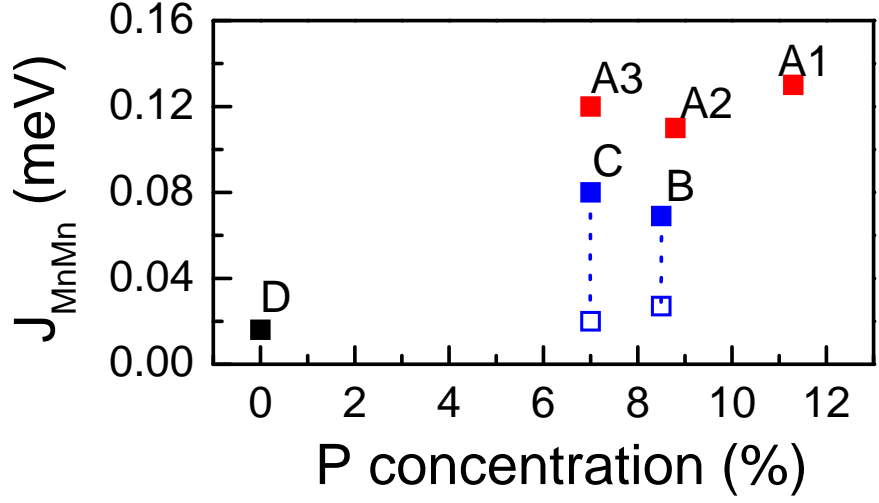} 
\caption{(Color online) Effective exchange energy between the Mn spins $J_{MnMn}$ as a function of P concentration at $T/T_C\approx 0.4$. For samples B and C the upper and lower boundaries for  $J_{MnMn}$ are shown by filled and open symbols, respectively.}
        \label{fig:Jmnmn}
\end{figure}

A direct comparison of the exchange constant $A$ between the A, B, C, and D samples is not relevant because of their different effective Mn concentrations. Therefore, using Kittel's relation between the first-neighbor spin-spin interaction constant and the exchange constant $A$ of the micromagnetic theory~\cite{Kittel1949}, we obtain an effective interaction constant between Manganese spins $J_{MnMn}$ independent of the Mn concentration as $J_{MnMn}=A a/2S^2$ where $a=(2/N_0 \left[Mn_{eff}\right])^{1/3}$ is the spin lattice parameter~\cite{note1}. 
Figure~\ref{fig:Jmnmn} represents $J_{MnMn}$ as a function of P concentration. The $J_{MnMn}$ values are shown for $T/T_C\approx0.4$, which is the lowest temperature for which data are available for all samples. $J_{MnMn}$ does not scale like $T_C$, which points to the importance of distant Mn interaction via the hole gas~\cite{Bouzerar2007}. $J_{MnMn}$ is larger for the P incorporated samples than for the (Ga,Mn)As sample by a factor up to 8. However, among the A series there is no clear tendency toward an increase of $J_{MnMn}$ with P concentration, at least within the narrow range investigated here. These samples with the largest Mn concentration (10.4\%) also have  the largest $J_{MnMn}$, which may result from a larger hole concentration. These results suggest a possible effect of Phosphorus incorporation on the Mn-hole exchange integral $J_{pd}$. A quantitative estimation of $J_{pd}$ is however not possible without the precise determination of the carrier concentration and a theoretical description of the hole density of states in the (Ga,Mn)As$_{1-y}$P$_y$ alloys, which is beyond the scope of this paper.
\par
\begin{figure}[t]
       \includegraphics[width=0.95\linewidth]{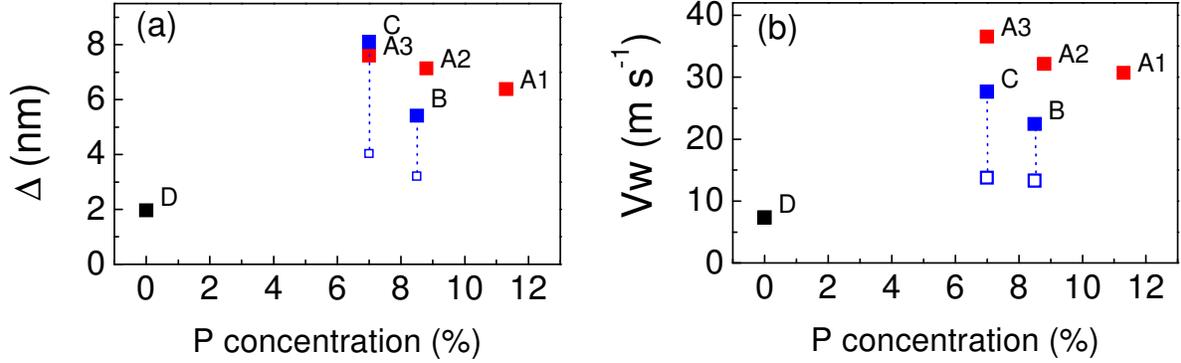} 
\caption{(Color online) DW width parameter $\Delta$ (a) and Walker velocity for DW propagation (b) as a function of P concentration at $T/T_C\approx0.4$. For samples B and C the upper and lower boundaries for these values are shown by filled and open symbols, respectively.}
      \label{fig:delta-vW}
\end{figure}

%
%\begin{figure}[t,b]
%        \includegraphics[width=0.7\linewidth]{Jmnmn.eps} 
%\caption{Effective exchange energy between the Mn spins $J_{MnMn}$ as a function of P concentration at $T/T_C\approx 0.4$. For samples B and C the upper and lower boundaries for  $J_{MnMn}$ are shown by filled and open symbols, respectively.}
%        \label{fig:Jmnmn}
%\end{figure}
%
Figure~\ref{fig:delta-vW}(a) shows the DW width parameter $\Delta$ as a function of P concentration. $\Delta$ is larger in P incorporated samples. This suggests that the DW mobility (field derivative of the DW velocity) in the field-driven DW propagation regimes, which scales like $\Delta$, should be larger in the P incorporated samples. We then obtain the Walker velocity $v_W=\gamma\Delta\mu_0 M_s /2$. We predict an increase of Walker velocity for the P incorporated samples, as shown in Fig.~\ref{fig:delta-vW}(b). DW propagation experiments are needed to confirm these predictions.

%\begin{figure}[t]
%        \includegraphics[width=0.95\linewidth]{Vw-delta.eps} 
%\caption{DW width parameter $\Delta$ (a) and Walker velocity for DW propagation (b) as a function of P concentration at $T/T_C\approx0.4$. For samples B and C the upper and lower boundaries for these values are shown by filled and open symbols, respectively.}
%        \label{fig:delta-vW}
%\end{figure}

%\section{conclusion}
\label{sec:conclusion}
As a conclusion, the novel ferromagnetic semiconductor (Ga,Mn)AsP is a very promising material for achieving FM layers with perpendicular easy axis. We have shown that the determination of the exchange constant and DW width parameter from domain theory is made more reliable than in (Ga,Mn)As owing to the self-organization of magnetic domains. This is made possible by the low density of DW pinning centers.

From the exchange constant $A$ an effective interaction constant between Mn spins $J_{MnMn}$ has been obtained. (Ga,Mn)As$_{1-y}$P$_y$ samples have a larger $J_{MnMn}$ than the (Ga,Mn)As/GaInAs sample, however no significant enhancement of $J_{MnMn}$ as a function of P concentration is found in the range investigated here. The Curie temperature and the ratio $[Mn_{eff}]/[Mn_{total}]$ do not exhibit any remarquable improvement.

The DW width parameter $\Delta$ as well as the predicted DW Walker velocity show a significant increase in the P incorporated samples. This should be of great importance for the study of field-driven and current-driven DW propagation.
 \\

This work was in parts supported by R\'{e}gion Ile de France under contract IF07-800/R with C'Nano IdF.

\newpage
\pagebreak[4]

%\begin{center}
%REFERENCES
%\end{center}

%
%
\newpage
\pagebreak[4]

\begin{table}[t]
\caption{\label{table1} (Ga,Mn)As$_{1-y}$P$_y$ and (Ga,Mn)As sample parameters. The lattice mismatch (lm) is defined as $(a_\bot-a_{sub})/a_{sub}$, where $a_\bot(a_{sub})$ is the lattice parameter of the film (substrate) parallel to the growth axis. $\epsilon_{zz}$ is the corresponding calculated strain component. $M_s$ and $Q$ are the magnetization and anisotropy quality factor, respectively, at temperature $T=4$~K. $Q$ slightly increases with temperature.\\}
\begin{ruledtabular}
\scriptsize
\begin{tabular*}{0.2\textwidth}{@{\extracolsep{2.5mm}}c|c|c|c||c||c||c}
sample&A1&A2&A3&B&C&D\\\hline
[Mn$_{total}$](\%)&10.4&10.4&10.4&8&7& 7 \\\
[P](\%)&11.3&8.8&7&8.5& 7&0 \\
lm (ppm)&-6650&-5120&-4260&-6870&-3580&-10700 \\
$\epsilon_{zz}$(\%)&-0.3&-0.2&-0.2&-0.3&-0.2&-0.5 \\
$T_C$~(K) &119&113&139&110&80&125 \\
$M_s$ (kA m$^{-1}$)&53.8&50.8&53.5&46.2&39.2&38.5\\
$Q$ &4.35&3.3&3&4.8&3.25&8.5 \\
\end{tabular*}
\end{ruledtabular}  
\end{table}

\newpage
\pagebreak[4]

\begin {center}
FIGURES
\end {center}


\end{document}